\documentclass[prb,aps,twocolumn,superscriptaddress,showpacs]{revtex4}
\usepackage{amsfonts,amsmath,bm}
\usepackage[OT4]{fontenc}

\usepackage{graphicx}

\newcommand{\ff}{\mathcal{F}}
\newcommand{\rr}{{\bm{r}}}
\newcommand{\kk}{\bm{q}}
\newcommand{\qq}{\bm{q}}
\newcommand{\kp}{\bm{k}\!\cdot\!\bm{p}}

\newcommand{\rl}{\rangle\!\langle}

\begin{document}

\title{Phonon-assisted relaxation and tunneling in self-assembled
  quantum dot molecules} 

\author{Krzysztof Gawarecki}
\affiliation{Institute of Physics, Wroc{\l}aw University of
Technology, 50-370 Wroc{\l}aw, Poland}
\author{Micha{\l} Pochwa{\l}a}
\affiliation{Department Physik, Fakult\"at f\"ur Naturwissenschaften and
  CeOPP, Universit\"at Paderborn, Warburger Strasse 100, D-33098
Paderborn, Germany}  
\author{Anna Grodecka--Grad}
\affiliation{QUANTOP, Danish National Research Foundation Center for
  Quantum Optics, Niels Bohr Institute, University of Copenhagen,
  DK-2100 Copenhagen \O, Denmark}  
\author{Pawe{\l} Machnikowski}
 \email{Pawel.Machnikowski@pwr.wroc.pl} 
\affiliation{Institute of Physics, Wroc{\l}aw University of
Technology, 50-370 Wroc{\l}aw, Poland}

\begin{abstract}
We study theoretically phonon-assisted relaxation processes in a
system consisting of one or two electrons confined in two vertically
stacked self-assembled quantum 
dots. The calculation is based on a $\bm{k}\!\cdot\!\bm{p}$
approximation for single particle 
wave functions in a strained self-assembled structure. From these,
two-particle states are calculated by including the Coulomb
interaction and 
the transition rates between the lowest energy eigenstates are derived.
We take into account phonon couplings via deformation potential and
piezoelectric interaction and show that they both can play a dominant
role in different parameter regimes. Within the Fermi golden rule
approximation, we calculate the relaxation rates between the
lowest energy eigenstates which lead to thermalization on a picosecond
time scale in a narrow range of dot sizes.
\end{abstract}


\maketitle

\section{Introduction} 
\label{sec:intro}

Structures composed of two closely spaced quantum dots (QDs) attract much 
attention motivated by their rich physical properties as well as by
possible applications in nanoelectronics or quantum computing. A major
factor that determines the properties of such a 
system is the electronic coupling between the dots. For closely spaced
dots, the system spectrum can be strongly affected by tunnel coupling 
\cite{bryant93,schliwa01,szafran01,bester04}. Optical spectra
of closely spaced structures indeed show clear manifestations of such
tunneling-related effects
\cite{ortner05,bayer01,ortner03,ortner05b,krenner05b}. Due to strong
delocalization of carrier states over the double dot structure,
analogous to a chemical covalent bond, such structures are
often referred to as quantum dot molecules (QDMs) or artificial molecules.

The properties of such artificial molecules are also
affected by phonon-related processes which are inevitable in a crystal
environment. Such effects will limit the feasibility of building
QDM-based quantum-coherent devices by providing a dephasing channel
for both charge
\cite{wu05,lopez05,stavrou05,vorojtsov05,climente06,grodecka08a,grodecka10} 
and spin
\cite{roszak09} states. Depending on the form and localization
character of the wave functions, such phonon-assisted transitions may either
take place between two delocalized states or involve charge
redistribution when an electron dissipatively tunnels to a different
dot. In the latter case, the electron spin can be conserved
\cite{kazimierczuk09}, which can be used to control the spin state of
a magnetic impurity in one of the QDs \cite{goryca09}.
Dissipative tunneling is also interesting in a two-electron
configuration, where a transition to a doubly occupied state is only
possible in a singlet configuration. This discrimination leads, on one
hand, to pure dephasing of singlet-triplet superpositions
\cite{roszak09} but, on the other hand, might be used to speed up the
proposed singlet-triplet measurement protocols \cite{barrett06}. 

From the experimental point of view, dissipative carrier transfer in
self-assembled structures has been studied with optical spectroscopy
methods (time-integrated and time-resolved photoluminescence, and
photoluminescence excitation experiments) both in lateral
double-dot systems \cite{rodt03,goryca09,kazimierczuk09}
as well as in stacked QDMs
\cite{heitz98,reischle07,nishibayashi08,park07,bajracharya07,sales04,%
mazur05,seufert01,tackeuchi00,ortner05c,chang08,gerardot05,nakaoka06}
and QD chains (both stacked and lateral) \cite{nakaoka04,wang08}. 
Various mechanisms have been invoked to account for the observed
properties. In most cases, the kinetics is attributed to
tunneling 
\cite{reischle07,nakaoka04,mazur05,seufert01,heitz98,tackeuchi00,%
rodt03,chang08,nakaoka06}. In some other experiments 
\cite{nishibayashi08,sales04,gerardot05}, signatures of radiative
(F\"orster-like) transfer have been observed. Coulomb scattering
\cite{ortner05c} and thermally activated processes
\cite{chang08,bajracharya07} also seem to play an important role, at
least in some systems. 

The variety of investigated structures and
probable transfer mechanisms is reflected in a relatively wide
distribution 
of measured transfer times. While the transfer in general takes place
on time scales shorter or comparable with the exciton life time (which
is necessary for the process to be observable in an optical
experiment), the observed times range from tens of picoseconds 
\cite{heitz98,nishibayashi08} to several nanoseconds
\cite{reischle07}. In most cases, however, transfer times between
hundreds of picoseconds and a few nanoseconds are observed. This
experimental situation indicates that carrier kinetics in coupled QD
structures is a rich and complex problem which most likely cannot be
solved by proposing a unique, universal theory. Therefore, it seems
reasonable to undertake a systematic theoretical study of various
carrier transfer processes and to identify conditions in which one or
another mechanism is expected to dominate the system properties. Such
a theoretical analysis of individual transfer mechanisms has in fact
already started with several works devoted to electron tunneling (in
simplified confinement models)  
\cite{wu05,lopez05,stavrou05,vorojtsov05,climente06,grodecka08a,%
grodecka10} and some studies of the F\"orster-like transfer
\cite{govorov03,govorov05,richter06,rozbicki08a}.

In this paper, we develop a theoretical description for
phonon-assisted relaxation and charge transfer (tunneling) in a
structure composed of two vertically stacked quantum dots formed in
the Stransky--Krastanov self-assembly process by strain-induced spontaneous
QD nucleation in the second layer on top of the
QDs formed in the first layer \cite{xie95,solomon96}. 
A reliable calculation of tunneling rates requires
reasonably precise knowledge of the electron wave functions. For a
strained self-assembled structure presently under consideration, this
implies the need to calculate the strain and then to find the single
particle wave functions, e.g., by a $\kp$ method. Then, Coulomb
interactions can be included for a two-electron system within the
standard configuration--interaction approach. 
The $\kp$ method for strained semiconductor heterostructures is a well
established procedure that has  been used for QDs, QDMs and other
nanostructures 
\cite{pryor98,pryor98b,korkusinski01,andrzejewski10}. Recently, this
method has been 
combined with the standard approach to carrier--phonon coupling in a
study of confined polarons \cite{obreschkow07}.
 Here, we apply a simplified
version of this method \cite{pochwala08} assuming a cylindrical
symmetry of the structure. This is motivated not only by economy of
computations but, more importantly, by the need to derive the wave
functions in a form suitable for efficient calculations of
carrier-phonon couplings and the following modeling of phonon-assisted 
relaxation.

The paper is organized as follows. In Sec.~\ref{sec:model}, we define
the model of the system under study. In Sec.~\ref{sec:strain}, we
discuss the strain fields in the structure. The one-electron and
two-electron states in the QDM are found in
Sec.~\ref{sec:single-electron} and Sec.~\ref{sec:coulomb},
respectively. In Sec.~\ref{sec:relax}, phonon-assisted relaxation for
one- and two-electron states is discussed. Concluding remarks and
discussion are contained in Sec.~\ref{sec:concl}.

\section{Model}
\label{sec:model}

\begin{figure}[tb]
\begin{center}
\includegraphics[width=55mm]{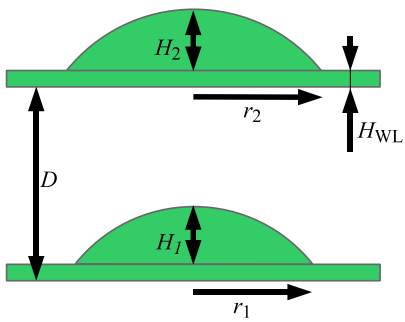}
\end{center}
\caption{\label{fig:struct}(Color online) The geometry of the QDM
structure.} 
\end{figure}

\begin{table*}
\begin{tabular}{llll}
\hline
& & GaAs & InAs \\
\hline
Lattice constant & $a$ & 0.56532~nm & 0.60583~nm \\
Elastic constants & $C_{11}$ &  $12.11\cdot10^{10}$~N/m$^{2}$ 
& $8.329\cdot10^{10}$~N/m$^{2}$ \\
 & $C_{12}$ &  $5.48\cdot10^{10}$~N/m$^{2}$ 
& $4.526\cdot10^{10}$~N/m$^{2}$ \\
 & $C_{44}$ &  $6.04\cdot10^{10}$~N/m$^{2}$ 
& $3.959\cdot10^{10}$~N/m$^{2}$ \\
Band structure parameters & $E_{\mathrm{c}0}$ & 0.95~eV & 0 \\
 & $E_{\mathrm{v}0}$ & -0.57~eV & -0.42~eV \\
 & $\Delta$ & 0.34~eV & 0.43~eV \\
 & $P_{0}$ & 9.89~eV & 9.19~eV \\
Deformation potentials & $a_{\mathrm{c}}$ & -9.3~eV & -6.66~eV \\
 & $a_{\mathrm{v}}$ & 0.7~eV & 0.66~eV \\
 & $b$ & -2.0~eV & -1.8~eV \\
Speed of sound & $c_{\mathrm{l}}$ & \multicolumn{2}{c}{5150~m/s} \\
& $c_{\mathrm{t}}$ & \multicolumn{2}{c}{2800~m/s} \\
Crystal density & $\varrho$ & \multicolumn{2}{c}{5300~kg/m$^{3}$} \\
Piezoelectric constant & $d_{\mathrm{P}}$ &
\multicolumn{2}{c}{-0.16~C/m$^{2}$} \\
Relative dielectric constant & $\varepsilon_{\mathrm{r}}$ &
\multicolumn{2}{c}{12.9} \\ 
\hline
\end{tabular}
\caption{\label{tab:param}System parameters used in the calculations.}
\end{table*}

We consider a QDM formed by two self-assembled InAs dots in a GaAs matrix.
The geometry of the structure as used in our modeling is shown in
Fig.~\ref{fig:struct}. 
The QDs are modeled as two spherical segments with base radii
$r_{1}$, $r_{2}$ and heights $H_{1}$, $H_{2}$, respectively.
Throughout the paper, the aspect ratio of the two dots will be held
constant, $H_{1}/r_{1}=H_{2}/r_{2}=0.37$.
Both dots
are placed on a wetting layer with thickness $H_{\mathrm{WL}}$. The
dots are separated by a distance $D$ (base to base). A 
diffusion layer of a very small thickness $H_{\mathrm{diff}}=0.3$~nm is
included at the contact between the two materials, 
in which the InAs concentration varies linearly. Apart from this, the
InAs content is assumed to be 100\% inside the dots and the wetting
layers and 0 outside. The parameters of the modeled structure are
collected in Table~\ref{tab:param}. 

Our model includes the case of a single electron in the QDM as well as
of two electrons coupled by the Coulomb interaction. The carriers
interact with bulk acoustic phonons via standard deformation potential
and piezoelectric interaction mechanisms.   

The modeling proceeds in three
steps: (1) Determination of the strain distribution; (2) Calculation
of the wave functions for single- and two-electron states; (3)
Calculation of relaxation rates. As each of these steps involves a
specific formalism, the corresponding details of the model will be
subsequently introduced in the following sections.

\section{Strain}
\label{sec:strain}

In this section, we calculate the strain present in the inhomogeneous
structure. 

The strain fields in the system will be described by the strain tensor 
\begin{displaymath}
\epsilon_{ij}(\bm{r})=\frac{1}{2}\left[
\frac{\partial u_{i}(\bm{r})}{\partial r_{j}} 
+\frac{\partial u_{j}(\bm{r})}{\partial r_{i}} \right], 
\end{displaymath}
where $\bm{u}(\bm{r})$ is the displacement field at the point $\bm{r}$
in the crystal. The elastic energy
of the inhomogeneous system is\cite{pryor98b}
\begin{eqnarray}
E_{\mathrm{el}} & = & \int d^{3}r \Big[ 
\frac{1}{2}C_{11}(\epsilon_{xx}^{2}+\epsilon_{yy}^{2}+\epsilon_{zz}^{2})
\nonumber \\
&&
+C_{12}(\epsilon_{yy}\epsilon_{zz}+\epsilon_{yy}\epsilon_{xx}
+\epsilon_{xx}\epsilon_{zz}) 
\nonumber \\
&&
+2C_{44}(\epsilon_{yz}^{2}+\epsilon_{zx}^{2}+\epsilon_{xy}^{2})
\nonumber \\
&&
-\alpha(\epsilon_{xx}+\epsilon_{yy}+\epsilon_{zz})\Big] .
\label{en-el}
\end{eqnarray}
Here $C_{ij}$ are
position-dependent elastic constants (see Table~\ref{tab:param} for the
values),  
$\alpha=(C_{11}+2C_{12})(a_{\mathrm{I}}/a_{\mathrm{G}}-1)$ in a QD and
$\alpha=0$ in GaAs, where
$a_{\mathrm{I}}$ and $a_{\mathrm{G}}$ are the lattice constants of
InAs and GaAs, respectively. The last term in Eq.~(\ref{en-el})
accounts for the mismatch of the lattice constants, shifting the
equilibrium of the InAs crystal lattice to the state appropriately
stretched with respect to the ideal InAs crystal. Since the
strain is calculated with respect to the GaAs lattice and GaAs
crystal coordinates are used, after minimizing the strain energy
the results for the InAs dots must be
rescaled to yield physical strain, according to\cite{pryor98b} 
\begin{equation}\label{rescale}
\epsilon_{ij}^{\mathrm{phy}}
=\frac{a_{\mathrm{G}}}{a_{\mathrm{I}}}\epsilon_{ij}
-\delta_{ij}\left(1-\frac{a_{\mathrm{G}}}{a_{\mathrm{I}}}\right).
\end{equation}

For an axially symmetric structure, it is convenient to perform the
computation in cylindrical coordinates $(\rho,\phi,z)$. Therefore we denote the
components of the displacement in the local reference frame as
$u_{\rho},u_{\phi},u_{z}$ and define the corresponding components of the
strain tensor, 
\begin{eqnarray*}
\epsilon_{\rho\rho} &= &\frac{\partial u_{\rho}}{\partial \rho},\\
\epsilon_{\phi\phi} &= &\frac{1}{\rho}\left(\frac{\partial u_{\phi}}{\partial \phi}
+u_{\rho} \right),\\
\epsilon_{\rho\phi} =  \epsilon_{\phi \rho}
&=&\frac{1}{2\rho}\left(\frac{\partial u_{\rho}}{\partial \phi}-u_{\phi}
\right),\\
\epsilon_{\rho z} =  \epsilon_{z\rho}
 & = & \frac{1}{2}\left( \frac{\partial u_{\rho}}{\partial z}+
\frac{\partial u_{z}}{\partial \rho} \right),\\
\epsilon_{\phi z} =\epsilon_{z\phi} & = &
\frac{1}{2}\left( \frac{\partial u_{\phi}}{\partial z}+
\frac{1}{\rho}\frac{\partial u_{z}}{\partial \phi} \right).
\end{eqnarray*}

We will look for the minimum of $E_{\mathrm{el}}$ in the class of
axially symmetric displacement fields, that is, $u_{\phi}=0$ and 
$\partial u_{r}/\partial\phi=\partial u_{z}/\partial\phi=0$.
With this assumption, the
integration over $\phi$ in Eq.~(\ref{en-el}) can be performed
analytically and one gets
\begin{eqnarray}
E_{\mathrm{el}} & = & \pi\int d\rho\rho\int dz\left[
C_{11}\epsilon_{zz}^{2}+D(\epsilon_{\rho\rho}^{2}+\epsilon_{\phi\phi}^{2})
+4C_{44}\epsilon_{\rho z}^{2}\right.\nonumber\\
&&+F\epsilon_{\rho\rho}\epsilon_{\phi\phi}
+2C_{12}\epsilon_{zz}(\epsilon_{\rho\rho}+\epsilon_{\phi\phi})
\nonumber\\
&&\left.-2\alpha(\epsilon_{\rho\rho}+\epsilon_{\phi\phi}+\epsilon_{zz})\right],
\label{en-el-cyl}
\end{eqnarray}
where $D=3C_{11}/4+C_{12}/4+C_{44}/2$ and
$F=C_{11}/4+3C_{12}/4-C_{44}/2$.

\begin{figure}[tb]
\begin{center}
\includegraphics[width=65mm]{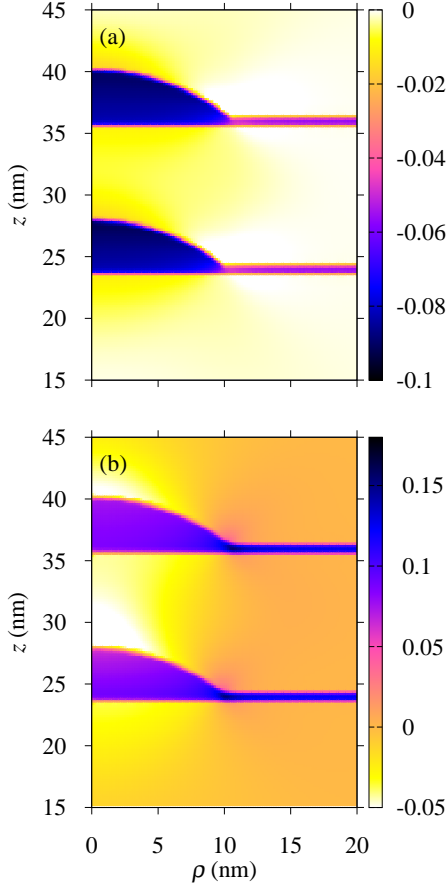}
\end{center}
\caption{\label{fig:strain}(Color online) The strain distribution in
  the structure for $D=12$~nm, $r_{1}=10$~nm, $r_{2}=10.5$~nm,
  $H_{1}=3.7$~nm, $H_{2}=3.885$~nm. In (a), the hydrostatic strain is shown; in
  (b), the axial strain
  $\epsilon_{zz}-(\epsilon_{\rho\rho}+\epsilon_{\phi\phi})/2$ is shown. } 
\end{figure}

The displacement field minimizing $E_{\mathrm{el}}$ is found by the conjugate
gradient method on a square grid of 1000 points along $z$ and 666
point along $\rho$, representing a cylinder with the height of $60$~nm
and the radius of $40$~nm. The boundary conditions represent
displacements due to the equilibrium strain in a system with two wetting
layers, which corresponds to the actual situation at large distances
from the dots. A combination of discretizations with 
forward and backward representations of derivatives is used to avoid
discretization-induced oscillations \cite{pryor98b}. 
As an example of the result, a strain map representing the hydrostatic
and axial strain across the structure for a 
selected geometry is shown in Fig.~\ref{fig:strain}.

\section{Single electron states}
\label{sec:single-electron}

In this section, we calculate approximate wave functions for a
single electron 
confined in the nanostructure. 
This is done within a variational multi-component envelope function
scheme \cite{pochwala08} based 
on the fact that the confinement volume is large compared to the
crystal lattice cell and that the local system parameters change
relatively slowly on atomic scales. 
In this approach, one finds the values of effective
masses and band edges at a given point by solving the bulk $\kp$ model
with strain and composition equal to those present at a given point.
This yields the band edge position, which is used as the local effective
potential, as well as the band curvatures, which define the components
of the effective mass tensor at a given point of the inhomogeneous
heterostructure. 

The conduction band structure in a strained system is determined from
the 8-band 
$\kp$ (Kane) Hamiltonian with strain-induced terms 
(Bir-Pikus Hamiltonian) using the L\"owdin elimination
\cite{lowdin51}. The part of the Hamiltonian coupling conduction and
valence band states is \cite{bahder90}
\begin{eqnarray*}
H_{\mathrm{c-v}} & = & |\mathrm{e}\!\uparrow\rangle\left[ 
-\sqrt{3}V^{\dag}\langle \mathrm{hh}\!\uparrow\! |
-U \left(\sqrt{2}\langle \mathrm{lh}\!\uparrow\! |
-\langle \mathrm{so}\!\uparrow\! |\right) \right.\\
&&\left.-V\left(\langle \mathrm{lh}\!\downarrow\! |
-\sqrt{2}\langle \mathrm{so}\!\downarrow\! |\right)
 \right] \\
&&
+|\mathrm{e}\!\downarrow\rangle\left[ 
-\sqrt{3}V\langle \mathrm{hh}\!\downarrow\! |
-U \left(\sqrt{2}\langle \mathrm{lh}\!\downarrow\! |
+\langle \mathrm{so}\!\downarrow\! |\right)
 \right.\\
&&\left.+V^{\dag}\left(\langle \mathrm{lh}\!\uparrow\! |
+\sqrt{2}\langle \mathrm{so}\!\uparrow\! |\right)
 \right] \Big.
+\mathrm{H.c.},
\end{eqnarray*}
where `e' `lh', `hh', and `so' denote the electron, heavy-hole, light-hole and
spin-orbit split-off
subbands, $\uparrow$ and $\downarrow$ represent the spin orientation
in a given subband, 
\begin{displaymath}
U=\frac{1}{\sqrt{3}}P_{0}\left(k_{z}+\sum_{j}\epsilon_{jz}k_{j}\right),
\end{displaymath}
and 
\begin{displaymath}
V=\frac{1}{\sqrt{6}}P_{0}
\left[k_{x}-ik_{y}-\sum_{j}(\epsilon_{xj}-\epsilon_{yj})k_{j}\right],
\end{displaymath}
where $P_{0}$ is proportional to the interband momentum matrix element (see
Table~\ref{tab:param} for parameter values).
The conduction band part of the Hamiltonian is
\begin{displaymath}
H_{\mathrm{c}}=
\left[ E_{\mathrm{c}0}+\frac{(\hbar k)^{2}}{2m_{0}}+a_{\mathrm{c}}h
 \right] 
(|\mathrm{e}\!\uparrow\rl
\mathrm{e}\!\uparrow\!|
+|\mathrm{e}\!\downarrow\rl \mathrm{e}\downarrow\!|),
\end{displaymath}
where $m_{0}$ is the free electron mass,
$E_{\mathrm{c}0}$ is the conduction band edge in a bulk unstrained
crystal, $a_{\mathrm{c}}$ is the conduction band deformation
potential, and $h=\mathrm{Tr}\epsilon$ is the 
hydrostatic strain.

As we are interested in the corrections to the conduction band
energies up to $k^{2}$ and the off-diagonal elements $U$ and $V$ are
proportional to $k$ we only need the conduction-valence band energy
difference at $k=0$. In this limit, the diagonal terms for the valence
band states are  
\begin{eqnarray*}
E_{\mathrm{hh}} & = & E_{\mathrm{v}0}-p-q, \\
E_{\mathrm{lh}} & = & E_{\mathrm{v}0}-p+q,\\
E_{\mathrm{so}} & = & E_{\mathrm{v}0}-\Delta -p,
\end{eqnarray*}
where $E_{\mathrm{v}0}$ is the valence band edge of an unstrained
crystal, $\Delta$ is the spin-orbit split-off parameter of an
unstrained crystal,
$p=a_{\mathrm{v}}h$, 
\begin{displaymath}
q=b\left[
\epsilon_{zz}-\frac{1}{2}(\epsilon_{\rho\rho}+\epsilon_{\phi\phi})\right],
\end{displaymath}
and $a_{\mathrm{v}},b$ are valence band deformation
potentials. The 
values of material parameters are given in Table~\ref{tab:param}.

Neglecting the strain-related terms in $U$ and $V$, which are much
smaller than the purely kinetic ones, we get the conduction band
energy up to the 2nd order in $\bm{k}$,
\begin{displaymath}
E(\bm{k})=E_{\mathrm{c}0}+a_{\mathrm{c}}h+\frac{\hbar^{2}k_{\bot}^{2}}{2m_{\bot}}
+\frac{\hbar^{2}k_{z}^{2}}{2m_{z}},
\end{displaymath}
where the in-plane and $z$ components
of the effective mass tensor are 
\begin{displaymath}
m_{\bot}^{-1}=m_{0}^{-1}\left( 
1+\frac{E_{\mathrm{P}}}{2\Delta E_{\mathrm{hh}}}
+\frac{E_{\mathrm{P}}}{6\Delta E_{\mathrm{lh}}}
+\frac{E_{\mathrm{P}}}{3\Delta E_{\mathrm{so}}} \right)
\end{displaymath}
and
\begin{displaymath}
m_{z}^{-1}=m_{0}^{-1}\left( 
1+\frac{2E_{\mathrm{P}}}{3\Delta E_{\mathrm{lh}}}
+\frac{E_{\mathrm{P}}}{3\Delta E_{\mathrm{so}}} \right),
\end{displaymath}
where $E_{\mathrm{P}}=2m_{0}P_{0}^{2}/\hbar^{2}$ and
$\Delta E_{i}=E_{\mathrm{c}0}+a_{\mathrm{c}}h-E_{i}$, for
$i=\mathrm{hh,lh,so}$. Note that $\Delta E_{i}$ are position dependent.

\begin{figure}[tb]
\begin{center}
\includegraphics[width=70mm]{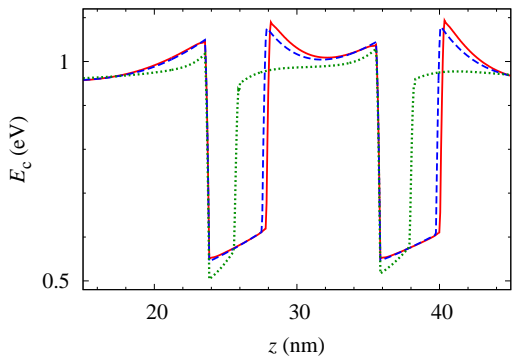}
\end{center}
\caption{\label{fig:bandedge}(Color online) The band edge profiles along
$z$ for three fixed values of $\rho=0.3$~nm (red solid line), $3$~nm (blue
dashed line), and $8$~nm (green dotted line) for the
structure as in Fig.~\ref{fig:strain}} 
\end{figure}

\begin{figure}[tb]
\begin{center}
\includegraphics[width=65mm]{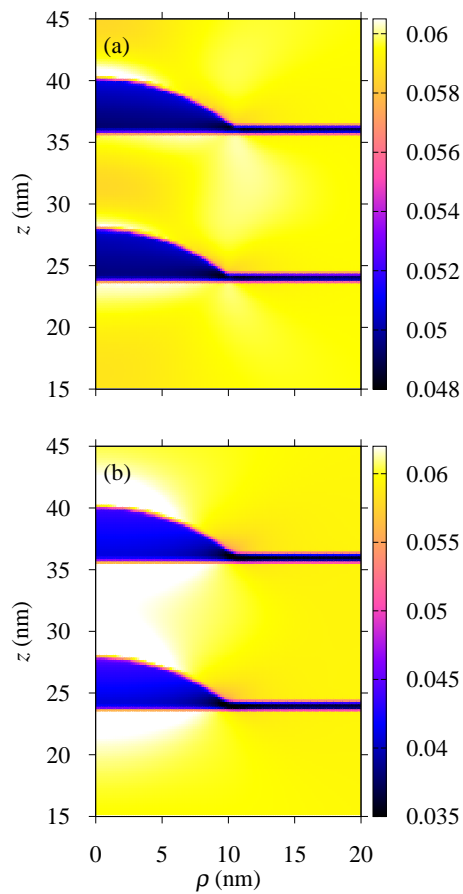}
\end{center}
\caption{\label{fig:masses}(Color online) The components of the
  electron effective mass for the structure as in
  Fig.~\ref{fig:strain}. (a) the radial component, $m_{\bot}$; 
  (b) the axial component, $m_{z}$.} 
\end{figure}

The dynamics of an electron in the strained
nanostructure in the present approach is defined by the 
conduction band edge at a given point,
$E_{\mathrm{c}}(\rho,z)=E_{\mathrm{c}0}+a_{\mathrm{c}}h(\rho,z)$ 
(which depends on the local strain),
and on the effective masses, which also vary across the structure. 
Fig.~\ref{fig:bandedge} shows an example of the  profiles of the conduction band edge
as a function of $z$ at three different values of $\rho$.
In Fig.~\ref{fig:masses}, we show the spatial maps of the radial and
axial components of the electron effective mass. Some strain-induced
anisotropy of the effective mass can be seen. The value of the radial
component, $m_{\bot}\approx 0.05m_{0}$ is close to the bulk GaAs value and
much higher than the bulk InAs value of $0.023m_{0}$.  It is roughly
constant within the volumes of the two dots. The
axial component $m_{z}$ is lower (about $0.04m_{0}$) and shows
some gradient along the QD axis, with higher values towards the top.

The envelope function of an electron is found from the Schr\"odinger
equation with the Hamiltonian
\begin{eqnarray*}
H & = & -\frac{\partial}{\partial x}
\frac{\hbar^{2}}{2m_{\bot}(\rho,z)} \frac{\partial}{\partial x}
-\frac{\partial}{\partial y}
\frac{\hbar^{2}}{2m_{\bot}(\rho,z)} \frac{\partial}{\partial y} \\
&&-\frac{\partial}{\partial z}
\frac{\hbar^{2}}{2m_{z}(\rho,z)} \frac{\partial}{\partial z}
+E_{\mathrm{c}}(\rho,z).
\end{eqnarray*}
Following the concept of `adiabatic' separation of variables
\cite{wojs96}, we first 
numerically solve the one-dimensional equation along the strongest
confinement direction at each $\rho$,
\begin{displaymath}
\left[  -\frac{\partial}{\partial z}
\frac{\hbar^{2}}{2m_{z}(\rho,z)} \frac{\partial}{\partial z}
+E_{\mathrm{c}}(\rho,z)\right]\chi(\rho,z)=E(\rho)\chi(\rho,z).
\end{displaymath}
The lowest two solutions to this equation, $\chi_{1,2}(\rho,z)$,
represent the lowest subband of confined states in the double well
system. The corresponding two branches of $\rho$-dependent eigenvalues,
$E_{1,2}(\rho)$ can be interpreted as effective potentials for the radial
problem. 

Next, we apply the Ritz variational method \cite{messiah66}, looking
for the stationary points of the functional 
\begin{displaymath}
F[\psi]=\langle \psi|H|\psi\rangle
\end{displaymath}
in the class of normalized
ansatz functions 
\begin{equation}\label{ansatz}
\psi(\rho,z,\phi)=\frac{1}{\sqrt{2\pi}}\sum_{\alpha}
\chi_{\alpha}(\rho,z)\varphi_{\alpha}(\rho)e^{iM\phi},
\end{equation}
where $M$ is the angular momentum.
Upon transforming to cylindrical coordinates and imposing the normalization
via the Lagrange multiplier $\lambda$, we write the functional
$F[\psi]$ in the explicit form
\begin{eqnarray*}
F[\psi] & = & \sum_{\alpha\beta}\int_{0}^{\infty}\rho d\rho\int_{-\infty}^{\infty}dz
\frac{\hbar^{2}}{2m_{\bot}(\rho,z)} \\
&&\times \frac{d}{d\rho}\left[ \chi_{\alpha}(\rho,z)\varphi_{\alpha}(\rho) \right]^{*} 
\frac{d}{d\rho}\left[ \chi_{\beta}(\rho,z)\varphi_{\beta}(\rho) \right] \\
&&+\sum_{\alpha}\int_{0}^{\infty}\rho d\rho \varphi_{\alpha}^{*}(\rho)
\left[ E_{\alpha}(\rho)+\frac{m^{2}}{\rho^{2}} \right] \varphi_{\alpha}(\rho) \\
&&-\lambda \left[  
\sum_{\alpha}\int_{0}^{\infty}\rho d\rho \varphi_{\alpha}^{*}(\rho)\varphi_{\alpha}(\rho)-1
\right].
\end{eqnarray*}
We discretize the functional $F[\psi]$ on the same lattice that
was used in the 
computation of the strain. As the functional is quadratic, the 
stationarity requirement with respect to the values at the discrete
points can easily be cast into the form of a matrix eigenvalue
problem for the components $(\varphi_{1},\varphi_{2})$. By virtue of the
Ritz theorem \cite{messiah66}, the corresponding eigenvalues
$\lambda_{n}$, $n=0,1,\ldots$
approximate the energy eigenvalues of the original
problem, while the eigenvectors, representing the components
$[\varphi_{1}^{(n)}(\rho),\varphi_{2}^{(n)}(\rho)]$ at the discrete
lattice points in the 
radial direction, are used to construct the electron eigenfunctions
$\psi_{n}(\bm{r})\equiv \psi_{n}(\rho,z,\phi)$
according to the ansatz formula (\ref{ansatz}). In this paper, the
discussion will be
restricted to the two lowest states for $M=0$, corresponding
to the tunnel-split ground state of the double-dot system.

\begin{figure}[tb]
\begin{center}
\includegraphics[width=85mm]{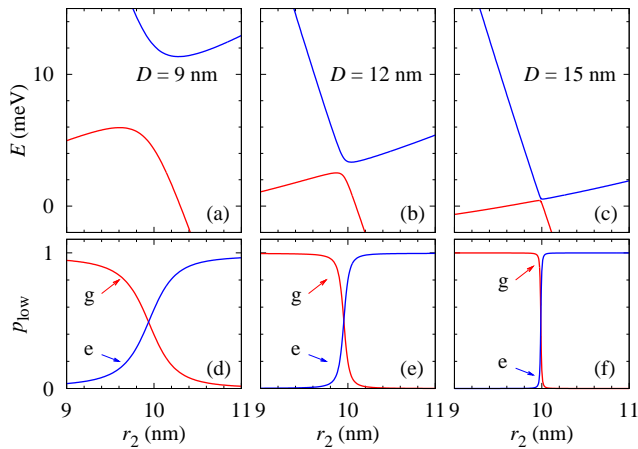}
\end{center}
\caption{\label{fig:one-electron}(Color online) (a-c) The single-electron
  energy levels for three structures with a fixed size of the
  lower dot as a function of the size (base radius $r_{2}$) of the
  upper one for three dot 
  separations as shown. Here $r_{1}=10$~nm, $H_{1}=3.7$~nm,
  $H_{2}/r_{2}=0.37$. The energy reference level is 0.8~eV above the
  conduction band edge of unstrained bulk InAs. (d-f) The corresponding
  probabilities 
  $p_{\mathrm{low}}$ of   finding the electron in the lower half of
  the system as a function of the size of the upper dot, in the ground
state (labelled ``g'') and in the first excited state (labelled ``e'')
of the system.}
\end{figure}

The single particle eigenenergies found within our approach for three
different distances $D$ between the dots are shown
in Fig.~\ref{fig:one-electron}(a-c). In these calculations, the shape
of the lower dot is kept constant, while the
base radius $r_{2}$ of the upper dot and its height $H_{2}$ are
varied, with $H_{2}/r_{2}$ constant. Electronic (tunnel) coupling
between the dots leads to the appearance of an anticrossing pattern
near the point where the dots become equal. The width of the
anticrossing is $2t$, where the phenomenological ``tunnel coupling''
$t$ is very well fitted by the formula $\ln t/t_{0}=-\kappa D$, with
$\kappa=0.59$~nm$^{-1}$ and $t_{0}=1.58$~eV. 
The value of $\kappa$ is consistent with the one-dimensional semiclassical
formula $\kappa=\sqrt{2m_{z}(V-E)}/\hbar$ if one uses $m_{z}=0.062m_{0}$ as
found in the area between the dots (see Fig.~\ref{fig:masses}), the
potential barrier height $V=1010$~meV (Fig.~\ref{fig:bandedge}) and
the electron energy $E=800$~meV (Fig.~\ref{fig:one-electron}).

In accordance with the spectral anticrossing, the electron occupations
for the ground and first excited states are transferred between the
dots, as shown in Figs.~\ref{fig:one-electron}(d-f). The exact
resonance point, where the 
two occupations are equal to 1/2 for both states (corresponding to
delocalized symmetric and antisymmetric wave functions), appears for
$r_{2}$ slightly smaller than $r_{1}$ which results from the strain
field in the absence of mirror symmetry in the structure. 

The procedure proposed here, involving the variational problem for a
two-component envelope, allows for mixing of the two manifolds of
states related to the two functions $\chi_{1}(\rho,z)$ and
$\chi_{2}(\rho,z)$, which is essential when the two QDs are of similar
size or when the thinner dot has a larger in-plane size, so that
a crossing of the one-dimensional solutions appears at a certain
value of $\rho$.

\section{Coulomb interaction and two-electron states}
\label{sec:coulomb}

We will find the two-electron states in the restricted basis of
low-energy configurations of the two electron system. We discuss the
situation when the energy
difference between the ground states in the two dots is smaller than
the intra-dot excitation energy (the latter is about 50~meV). Then the
two lowest single-particle states found in
Sec.~\ref{sec:single-electron} correspond to an electron in the ground
state of one of the dots or, near the resonance, to a delocalized
superposition of the two ground states. 

Let $a_{n,\sigma},a_{n,\sigma}^{\dag}$ denote the 
annihilation and creation
operators for an electron in the state $n=0,1$ with the
wave function $\psi_{n}(\bm{r})$ and spin $\sigma$. The low-energy two-electron
configurations split into one triplet state (of no interest in the
present discussion) and three singlet states
\begin{subequations}
\begin{eqnarray}
\label{2-el-basis}
|0 \rangle & = &
a_{0\uparrow}^{\dag}a_{0\downarrow}^{\dag} |\mathrm{vac}
\rangle, \\ 
|1 \rangle & = & 
\frac{a_{0\uparrow}^{\dag}a_{1\downarrow}^{\dag}
+a_{1\uparrow}^{\dag}a_{0\downarrow}^{\dag}}{\sqrt2}
|\mathrm{vac} \rangle,\\ 
|2\rangle & = &
a_{1\uparrow}^{\dag}a_{1\downarrow}^{\dag} |\mathrm{vac}
\rangle,
\end{eqnarray}
\end{subequations}
where $|\mathrm{vac} \rangle$ is the vacuum (empty dot) state.

The Hamiltonian of the interacting two-electron system has the form 
\begin{equation}\label{ham-coulomb}
H =  \sum_{n,\sigma} \epsilon_{n} a_{n\sigma}^{\dag}a_{n\sigma}
+\frac{1}{2} \sum_{ijkl} \sum_{\sigma,\sigma'} 
v_{ijkl}
a_{i\sigma}^{\dag}a_{j\sigma'}^{\dag}  a_{k\sigma'} a_{l\sigma},
\end{equation}
where 
\begin{eqnarray}
v_{ijkl} & = & 
\frac{e^{2}}{4 \pi \varepsilon_{0} \varepsilon_{\mathrm{r}}} \int d^{3}\bm{r}
\int d^{3}\bm{r}' \nonumber \\
&&\times \psi_{i}^{*}(\bm{r}) \psi_{j}^{*}(\bm{r}') 
\frac{1}{| \bm{r}-\bm{r}'|}
\psi_{k}(\bm{r}') \psi_{l}(\bm{r}). \label{v-ijkl}
\end{eqnarray}
Here $e$ is the electron charge, $\varepsilon_{0}$ is the vacuum
permittivity, and $\varepsilon_{\mathrm{r}}$ is the dielectric constant of
the semiconductor. Some technical details concerning the calculation
of Coulomb matrix elements for the wave functions obtained within the
variational two-component envelope function scheme in
Sec.~\ref{sec:single-electron} are given in the Appendix.

\begin{figure}[tb]
\begin{center}
\includegraphics[width=85mm]{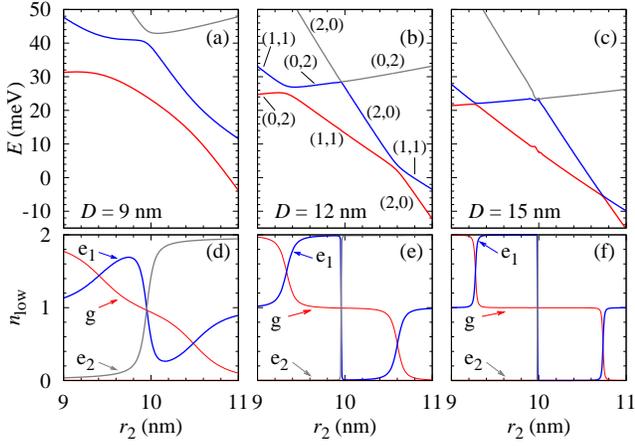}
\end{center}
\caption{\label{fig:two-electron}(Color online) (a-c) The two-electron
  energy levels for three structures with a fixed size of the
  lower dot as a function of the size (base radius $r_{2}$) of the
  upper one for three dot 
  separations as shown. Here $r_{1}=10$~nm, $H_{1}=3.7$~nm,
  $H_{2}/r_{2}=0.37$. The energy reference level is 1.6~eV above the
  conduction band edge of unstrained bulk InAs. In (b), the electron
  configurations corresponding to the spectral branches are shown,
  with the first and second digits corresponding to the number of
  electrons in the upper and lower dot, respectively.  (d-f) The
  corresponding 
  average numbers of electrons in the lower half of
  the system as a function of the size of the upper dot, in the ground
state (labelled ``g'') and in the first and second excited states
(labelled ``e1,e2'') of the system.}
\end{figure}

In Fig.~\ref{fig:two-electron}, we show the three lowest spin-singlet
eigenstates of the interacting 
two-electron system as a function of the size of the upper dot with
the lower dot kept fixed. The central resonance occurs when the dots
are close to identical and involves the doubly occupied configurations
$(0,2)$ and $(2,0)$ which, at the resonance point,
have similar energy. This anticrossing is very narrow (less than
$0.1$~meV for $D=12$~nm) since the two states involved differ by the
location of both electrons [see Figs.~\ref{fig:two-electron}(d)-(f)]
and, therefore, are coupled only by very 
small exchange-like Coulomb terms. Only for the smallest inter-dot
distance considered, $D=9$~nm, this splitting becomes larger due to
stronger mixing of configurations and incomplete electron localization
in the two states (which allows the configurations to be coupled by
single-electron tunneling). The other two anticrossings occur at the
degeneracy point between the singly occupied $(1,1)$
configuration (favored by the Coulomb repulsion) and the
$(0,2)$ or $(2,0)$ configuration with two electrons in
the larger dot. One can notice that these two splittings are wider
than those appearing between the single-electron states, shown in
Fig.~\ref{fig:one-electron} (for instance, 2~meV vs. 1.5~meV for
$D=12$~nm). This is due to the fact that the anticrossing of
two-electron configurations is enhanced by Coulomb terms
\cite{grodecka08a}.  

\section{Phonon-assisted relaxation}
\label{sec:relax}

In this section, we discuss the phonon-assisted relaxation between the
single-electron states and between the two lowest two-electron
states. 

The coupling between the electrons and phonons is described by the
Hamiltonian
\begin{equation}\label{ham-e-ph}
H_{\mathrm{e-ph}}=\sum_{nm,\sigma} a_{n,\sigma}^{\dag}
a_{m,\sigma}^{\phantom{\dag}}
\sum_{s,\kk} F_{s,nm}(\kk) 
\left(b_{s,\kk}^{\phantom{\dag}} + b_{s,-\kk}^{\dag} \right),
\end{equation}
where the coupling constants $F_{s,nn'}(\kk)$ 
have the symmetry $F_{s,nn'}(\kk) = F_{s,n'n}^{*}(-\kk)$. 
The inter-level energy distance in our structure is smaller than
the optical phonon energy. Therefore, only acoustic phonons are relevant in our
model. We include the deformation potential (DP) coupling to longitudinal
acoustic (LA) phonons and the piezoelectric (PE) coupling to LA as well as
transverse acoustic (TA) phonons. The coupling constant for the DP
coupling mechanism is given by 
\begin{equation}\label{FDP}
F^{\rm DP}_{{\rm l},nn'}(\kk) = \sqrt{\frac{\hbar q}{2 \rho V c_{\rm
l}}} a_{\mathrm{v}} \ff_{nn'} (\kk),
\end{equation}
where $\varrho$ is the crystal density, $V$ is the normalization volume
of the phonon modes, $c_{\rm l}$ is the longitudinal speed of sound
(see Table~\ref{tab:param} for parameter values),
and the form factor is defined as 
\begin{equation}\label{FF}
\ff_{nn'}=\int d^{3}r 
\psi^{*}_{n}(\rr)e^{i\kk\cdot\rr} \psi_{n'}(\rr).
\end{equation}
The coupling element for PE interactions reads
\begin{equation}\label{FPE}
F^{\rm PE}_{s,nn'}(\kk) = -i \sqrt{\frac{\hbar}{2 \rho V c_s q}}
\frac{d_{\rm P} e}{\varepsilon_{0}\varepsilon_{\rm r}} M_s (\hat\kk) \ff_{nn'} (\kk),
\end{equation}
where $c_s$ is the speed of sound ($s=\,$l,t denotes the LA and TA
phonon branch, respectively) and $d_{\rm
P}$ is the piezoelectric constant. The function $M_s (\hat\kk)$ does
not depend on the value of the phonon wave vector, but only on its
orientation. For a zinc-blende structure, it reads
\begin{eqnarray} \nonumber
M_s (\hat\kk) & = & \hat q_{x}
\left[ (\hat e_{s,\kk})_{y}\hat q_{z} 
+ (\hat e_{s,\kk})_{z}\hat q_{y} \right] \\ \nonumber
&& + \hat q_{y} \left[ (\hat e_{s,\kk})_{z}\hat q_{x} + (\hat
e_{s,\kk})_{x}\hat q_{z} \right] \\ && + \hat q_{z} \left[ (\hat
e_{s,\kk})_{x}\hat q_{y} + (\hat e_{s,\kk})_{y}\hat q_{x} \right],
\end{eqnarray}
where $\hat e_{s,\kk}$ is the unit polarization vector for the phonon
wave vector $\kk$ and polarization $s$, and $\hat\kk = \kk/q$.  We
choose the following phonon polarization vectors
\begin{eqnarray}
\hat e_{{\rm l},\kk} & \equiv & \hat\kk = \left( 
\sin \theta \cos\phi, \sin \theta \sin \phi, \cos \theta \right),\\ \nonumber
\hat e_{{\rm t1},\kk} & = & \left( 
-\sin \phi, \cos \phi ,0 \right),\\ \nonumber
\hat e_{{\rm t2},\kk} & = & \left( 
\cos \theta\cos\phi, \cos \theta \sin \phi, -\sin \theta \right),
\end{eqnarray}
for which the functions $M_{s}(\hat\kk)$ read
\begin{eqnarray}\label{Ms}
M_{\rm l} (\hat\kk) & = & \frac{3}{2} \sin(2\theta)\sin\theta
\sin(2\phi), \\ 
\nonumber M_{\rm t1} (\hat\kk) & = & 
\sin(2\theta) \cos(2\phi), \\ \nonumber M_{\rm t2} (\hat\kk) & = &
\sin\theta \left(3 \cos^{2} \theta -1 \right) \sin(2\phi).
\end{eqnarray}

In what follows, we will assume that higher states are separated by an
energy much 
larger than $k_{\mathrm{B}}T$, where $k_{\mathrm{B}}$ is the Boltzmann
constant and $T$ is the temperature. 
Then, the kinetics
leading to thermalization of the occupations of the two relevant
levels can be characterized by the occupation of the upper state,
\begin{displaymath}
n(t)-n_{\mathrm{eq}}=\left( n_{0}-n_{\mathrm{eq}} \right) e^{-\gamma t},
\end{displaymath}
where $n_{0}$ is the initial occupation, $\gamma$
is the relaxation (thermalization) rate and 
\begin{displaymath}
n_{\mathrm{eq}}=\frac{1}{e^{\Delta E/(k_{\mathrm{B}}T)}+1}
\end{displaymath}
is the equilibrium occupation, where $\Delta E>0$ is the energy
separation between the two states. 

Thus, given the initial condition and the energy difference $\Delta
E$, the thermalization kinetics is determined by the relaxation rate
$\gamma$ (or the relaxation time $\tau=\gamma^{-1}$) which will be found in
the following sections, first for a single-electron, then for the
two-electron case. 

\subsection{Single electron relaxation}
\label{sec:relax-single}

For a single electron system, the thermalization rate $\gamma$ can be
found directly from Eq.~(\ref{ham-e-ph}) using the Fermi golden
rule. The result can be written in the form  
\begin{equation}\label{FGR}
\gamma=2\pi\left[2n_{\mathrm{B}}(\Delta E)+1\right] J(\Delta E/\hbar),
\end{equation}
where 
\begin{displaymath}
n_{\mathrm{B}}(\Delta E)=\frac{1}{e^{\Delta E/(k_{\mathrm{B}}T)}-1}
\end{displaymath}
is the Bose distribution and the spectral density $J(\omega)$ is given
by
\begin{equation}\label{spdens}
J(\omega)=\frac{1}{\hbar^{2}}\sum_{\kk,s}
|F_{s,01}(\kk)|^{2}\delta(\omega-\omega_{\kk,s}),
\end{equation}
where $F_{s,01}(\kk)$ is the total coupling for the branch $s$, that
is, $F_{\mathrm{l},01}(\kk)=
F_{\mathrm{l},01}^{(\mathrm{PE})}(\kk)+F_{\mathrm{l},01}^{(\mathrm{DP})}(\kk)$
and
$F_{\mathrm{s},01}(\kk)=F_{\mathrm{t},01}^{(\mathrm{PE})}(\kk)$ for
$s=\mathrm{t1,t2}$. In fact, due to different parity of the DP and PE
couplings (as functions of $\kk$) the two contributions do not
interfere and the spectral density (hence, also the thermalization
rate) can be split into the corresponding two parts
$J^{(\mathrm{DP})}(\omega)$ and $J^{(\mathrm{PE})}(\omega)$.

In order to find the thermalization rate, we calculate the
form-factors defined in Eq.~(\ref{FF}) using the single-electron wave
functions found for the strained double dot structure in
Sec.~\ref{sec:single-electron} (see Appendix). From these,
we find the coupling 
constants given by Eq.~(\ref{FDP}) and Eq.~(\ref{FPE}) and the
corresponding spectral densities given by Eq.~(\ref{spdens}). The rate
$\gamma$ then follows from the Fermi golden rule formula,
Eq.~(\ref{FGR}). 

\begin{figure}[tb]
\begin{center}
\includegraphics[width=85mm]{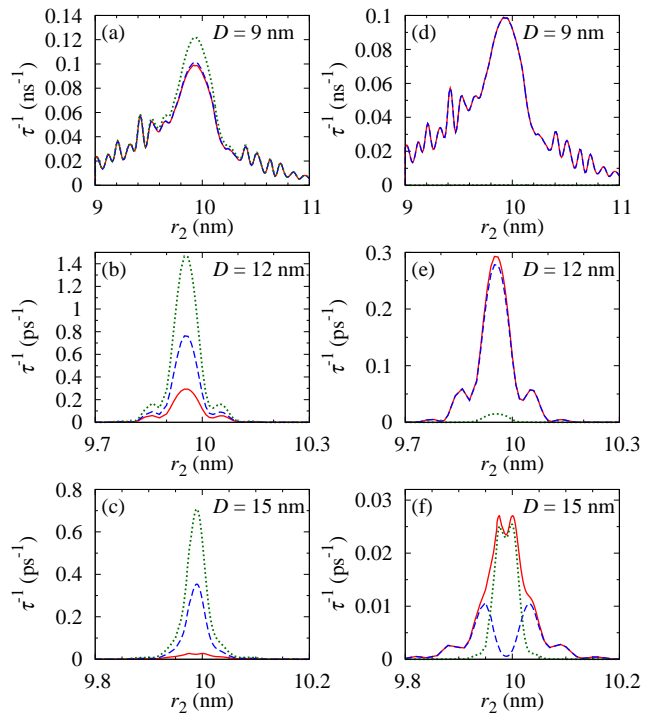}
\end{center}
\caption{\label{fig:tun-one-electron}(Color online) (a-c)
  Thermalization rate for one electron states in a structure with
  $r_{1}=10$~nm, $H_{1}=3.7$~nm and $H_{2}/r_{2}=0.37$ for three different
  inter-dot separations at $T=0$~K (red solid
  line), 20~K (blue dashed line), and 40~K (green dotted line). (d-f)
  Contributions to the thermalization rate from the DP
  coupling (blue dashed line) and PE coupling
  (green dotted line) as well as the total rate (red solid line) at
  $T=0$~K.} 
\end{figure}

\begin{figure}[tb]
\begin{center}
\includegraphics[width=85mm]{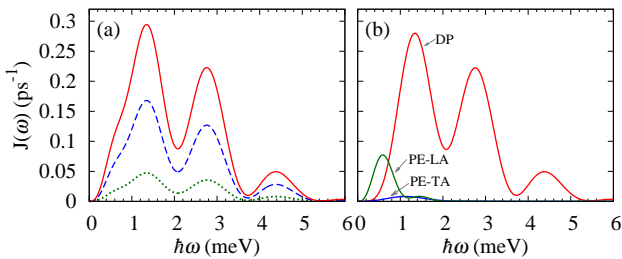}
\end{center}
\caption{\label{fig:spdens}(Color online) (a) The spectral density of the
  phonon reservoir as a function of frequency for three different
  values of the upper dot radius in the resonance area. red solid line:
  $r_{2}=9.96$~nm (exact resonance); blue dashed line: $r_{2}=9.9$~nm;
  green dotted line: $r_{2}=97.8$~nm.  (b) The contributions to the
  spectral density at the resonance from the deformation potential
  coupling to LA phonons and from the piezoelectric coupling to LA and
  TA phonons.} 
\end{figure}

The single particle relaxation rates are shown in
Figs.~\ref{fig:tun-one-electron}(a-c) as functions of 
the upper dot 
size for three values of the inter-dot spacing (for the same sample
geometries as in Fig.~\ref{fig:one-electron}) and at three different
temperatures. These three plots show that both the magnitude and the
size dependence of the relaxation rate is different in these three
cases. The interpretation of this behavior can be based on the Fermi
golden rule in the form of Eq.~(\ref{FGR}), where the essential role
is played by the spectral density defined in Eq.~(\ref{spdens}) and
plotted (for $D=12$~nm) in Fig.~\ref{fig:spdens}. 

The overall magnitude of the
spectral density depends on the spatial overlap between the wave
functions corresponding to the states involved in the
transition. It is, therefore, large at the resonance and becomes
smaller as the system is shifted off the resonance point 
Fig.~\ref{fig:spdens}(a). Apart from
this, the spectral density shows oscillations in its high-frequency
tail which are due to the essentially one-dimensional emission of
short wavelength phonons along the strongest confinement direction
\cite{bockelmann90}. As we deal with two confinement centers displaced
along the same direction, interference effects appear and the phonon
emission amplitude has a maximum whenever $\omega=(2n+1)\pi c/D$ for
an integer $n$. Moreover, the envelope of the spectral density decays
at high frequencies since the short wave length phonons are not
effectively coupled to the relatively weakly confined electron states.

In the case of closely stacked dots
[Fig.~\ref{fig:tun-one-electron}(a)], the tunnel splitting of the QDM
electron states is large and the
frequency of the emitted phonons always lies far in the tail of the
spectral density. This is reflected by the very low relaxation
rate. The oscillations of the spectral density are clearly marked in
the values of the relaxation rate.
When the dots are separated by a larger distance
[Fig.~\ref{fig:tun-one-electron}(b)], the resonance
becomes narrower and now the resonant frequency lies in the region of
large spectral density. When moving away from the resonance, the
relaxation rate drops down primarily due to the decreasing overlap of
the wave functions. This leads to a narrow peak in the dependence of
the relaxation rate around the resonance value.  Still, oscillations
are visible in the slopes of this peak.
For even larger inter-dot distances
[Fig.~\ref{fig:tun-one-electron}(c)], the resonance becomes very 
narrow. Correspondingly, the overlap between the wave functions decays
almost completely already when the size of the upper dot is changed by
a fraction of a nanometer from the resonant value. Therefore,
the relaxation rate is large only in a very narrow region around the
resonance. The rates are also generally lower than in the previous
case, which results from the dependence of the spectral density at low
frequencies ($\sim \omega^{5}$ for the DP coupling
and $\sim \omega^{3}$ for the PE coupling). 

The interplay between the shape and magnitude of the spectral densities
for different coupling mechanisms [Fig.~\ref{fig:spdens}(b)] and the
electron energies near the resonance is reflected also by the
different contributions from the DP and PE couplings to the total
relaxation rates. As can be seen in
Figs.~\ref{fig:tun-one-electron}(d-f), the DP coupling dominates for
large energy splittings. The reason is that this coupling is isotropic
and involves LA phonons which have higher energies. On the contrary,
the piezoelectric coupling is anisotropic and, according to
Eq.~(\ref{Ms}), is suppressed for emission along the $z$ direction
that is preferred at high frequencies. The situation changes at low
energy splittings where the low-frequency properties of the spectral
density are relevant. As the spectral density for the piezoelectric
coupling decreases at $\omega\to 0$ more slowly than that
corresponding to the DP coupling the PE coupling is the dominating
mechanism in the case of narrow anticrossing, as can be seen in 
Fig.~\ref{fig:tun-one-electron}(f). 
For very low frequencies, all the contributions to the spectral density
are small, hence the phonon-assisted relaxation process becomes
ineffective for small energy splitting. This is manifested by a dip in
the thermalization rate at the exact resonance for $D=12$~nm
[Fig.~\ref{fig:tun-one-electron}(f)]. 

\begin{figure}[tb]
\begin{center}
\includegraphics[width=85mm]{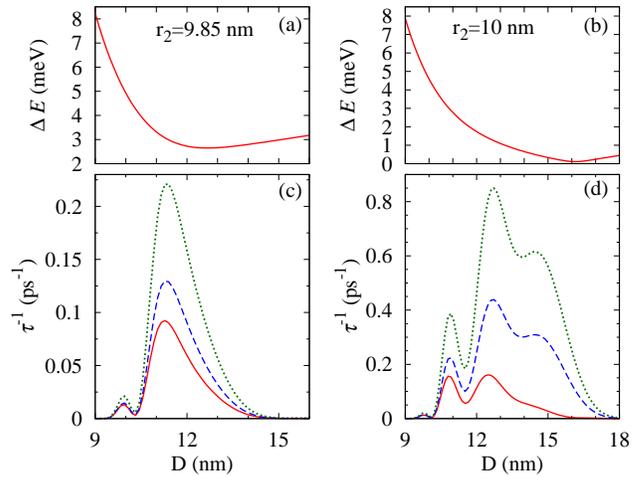}
\end{center}
\caption{\label{fig:rate-d}(Color online) (a,b) Energy splitting
  between the two lowest single-electron states as a function of the
  distance $D$ between the dots for $r_{1}=10$~nm and $r_{2}$ as
  shown. (c,d) Thermalization rate for one electron states in a structure with
  $r_{1}=10$~nm, $H_{1}=3.7$~nm and $H_{2}/r_{2}=0.37$ as a function
  of $D$ for the two values of $r_{2}$  at $T=0$~K (red solid
  line), 20~K (blue dashed line), and 40~K (green dotted line).} 
\end{figure}

In Fig.~\ref{fig:rate-d} we show the energy splitting between the two
lowest energy levels and the corresponding values of the
thermalization rates $\gamma=\tau^{-1}$ as a function of the inter-dot
distance $D$ for two system geometries: slightly different dots
[Fig.~\ref{fig:rate-d}(a,c)] and identical dots
[Fig.~\ref{fig:rate-d}(b,d)]. The values of the rates show
oscillations, resulting from the variation of the energy level
splitting and corresponding to the oscillations of the spectral
density, as discussed above. The maximum value is quite large and
corresponds to relaxation times of several picoseconds, which results
from the relative proximity of the resonance (identical dots) in both
presented cases. The maximum goes down and shifts to lower distances
as the dots become different. At large distances the relaxation
becomes inefficient in any case. In an attempt (not shown) to compare the decrease
of the rates at large $D$ with an exponential law (as observed, at
least approximately, in some experiments
\cite{heitz98,tackeuchi00,mazur05,reischle07}), we have found a roughly
exponential decay with a coefficient consistent with
the value of $\kappa$ found in Sec.~\ref{sec:single-electron}. This
decay is, however, strongly modulated by oscillations. This results
from a small energy scales in our model which is comparable to
strain-related effects as the dots are moved with respect to each
other. This is visible in Figs.~\ref{fig:rate-d}(a,c), where the
energy level separation does not tend to a constant asymptotic value at
large $D$ as would be expected for a simple model of two
potential wells with fixed shapes.

\subsection{Relaxation in two-electron systems}
\label{sec:relax-two}

In this section, we calculate the transition rates for phonon-assisted relaxation
between two-electron states $|\Psi_{i}\rangle$, obtained from the
diagonalization of the Hamiltonian (\ref{ham-coulomb}) in the
restricted basis formed by the states $|0\rangle,|1\rangle$ and $|2\rangle$
[Eqs.~(\ref{2-el-basis})-(c)]. We first project the carrier-phonon
Hamiltonian (\ref{ham-e-ph}) onto the two-electron subspace,
\begin{displaymath}
H_{\mathrm{e-ph}}^{(2)}=\sum_{ij}|\Psi_{i}\rl\Psi_{j}|
\sum_{s,\kk}G_{s,ij}(\kk)
\left(b_{s,\kk}^{\phantom{\dag}} + b_{s,-\kk}^{\dag} \right),
\end{displaymath}
where the coupling constants
\begin{displaymath}
G_{s,ij}(\kk)=\sum_{nm,\sigma}\left\langle\Psi_{i}
|a_{n,\sigma}^{\dag}a_{m,\sigma}|\Psi_{j} \right\rangle F_{s,nm}(\kk)
\end{displaymath}
are found based on the numerical results for the states
$|\Psi_{i}\rangle$. 
We restrict the discussion to transitions between the two lowest
states $|\Psi_{0}\rangle$ and $|\Psi_{1}\rangle$, separated by an
energy splitting $\Delta E$. 
In the Fermi golden rule approximation, the rate for the
relaxation of the two occupations to equilibrium is
\begin{equation*}
\gamma^{(2)}=2\pi\left[2n_{\mathrm{B}}(\Delta E)+1\right]
J^{(2)}(\Delta E/\hbar),
\end{equation*}
where the spectral density $J^{(2)}(\omega)$ is given
by
\begin{equation*}
J^{(2)}(\omega)=\frac{1}{\hbar^{2}}\sum_{\kk,s}
|G_{s,01}(\kk)|^{2}\delta(\omega-\omega_{\kk,s}).
\end{equation*}

\begin{figure}[tb]
\begin{center}
\includegraphics[width=85mm]{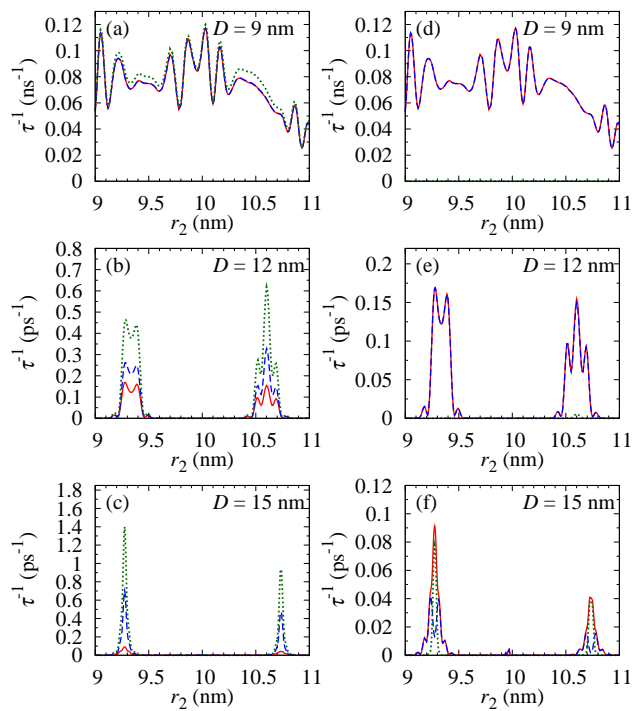}
\end{center}
\caption{\label{fig:tun-two-electron}(Color online) (a-c)
  Thermalization rate between the two lowest two-electron states for
  three different 
  inter-dot separations at $T=0$~K (red solid
  line), 20~K (blue dashed line), and 40~K (green dotted line). (d-f)
  Contributions to the thermalization rate from the DP coupling (blue
  dashed line) and PE coupling
  (green dotted line) as well as the total rate (red solid line) at
  $T=0$~K.} 
\end{figure}

The inverse relaxation times $\tau^{-1}=\gamma^{(2)}$ resulting from
these calculations are presented in 
Fig.~\ref{fig:tun-two-electron}. Like in the single electron case, the
energy level splitting in the case of relatively closely spaced dots
($D=9$~nm) is very large and the resonance is very broad which results
in very long relaxation times which do not vary considerably over the
parameter range studied [Fig.~\ref{fig:tun-two-electron}(a)]. For such
high transition energies, only LA phonons contribute to the process
via DP coupling  [Fig.~\ref{fig:tun-two-electron}(d)]. At larger
inter-dot distances, the transition rates become large around the
resonance points corresponding to the anticrossing of $(1,1)$ and
$(2,0)$ or $(0,2)$ configurations. The structure of the relaxation
rate as a function of the upper dot diameter $r_{2}$ is similar to
that discussed in the single electron case above. Also the relative
contributions form different coupling mechanisms behave in the same
way, with the piezoelectric coupling dominating at low energies. In
general, the relaxation rates are similar to those found in the single
electron case, since both these processes are physically very similar.
In both cases, the electron tunnels between the dots and simultaneously
emits one phonon. The only difference is that  
in the single electron case it tunnels towards an empty QD, whereas
in the two electron case, there is already another electron. 
This basically leads to shifts (due to Coulomb interaction) of the
parameter regimes where the relaxation is most efficient from the
region of identical dots to the asymmetric situation where the
difference of confinement energies compensates for the Coulomb
repulsion.  
A similar conclusion has been reached in the case of gated QDM
structures modeled by 
Gaussian potential wells \cite{grodecka08a}.  

\section{Conclusion}
\label{sec:concl}

We have studied phonon-assisted relaxation (thermalization) for
single-electron and two-electron configurations in self-assembled
quantum dots. In order to describe the electron states in a strained
structure in a possibly 
realistic (but still relatively simple) way and to reliably model the
effect of the system geometry we have developed a
generalized, multi-component envelope function formalism based on the
variational principle. 

Our results show that the single phonon relaxation is very efficient
in an extremely
narrow range of relative QD sizes near the anticrossings of energy
levels but only for systems with a sufficiently large inter-dot
distance (several nm). In this case, the relaxation times can be as
low as 1~ps, both in the single-electron and two-electron cases. The
range of efficient relaxation becomes narrower as the dots are more
distant from each other. 
Both coupling channels, piezoelectric and
deformation potential, are important for the overall relaxation
rate. The former dominates at low (sub-meV) transition energies.

When
the distance between the dots becomes smaller than about 10~nm, the 
energy level splitting becomes too large to be spanned by a single
acoustic phonon (but still to small for an optical one). In this range
of closely stacked dots, the tunneling times increase by orders of
magnitudes and take values in the nanosecond range. For such small
inter-dot distances, the energy splitting between the two lowest
states is dominated by tunnel coupling and depends weakly on the size
difference. As a result, the efficiency of the relaxation process
remains nearly constant over a wide range of dot sizes. One can
expect, however, that two-phonon processes \cite{stavrou06} can be
important in this range of parameters, in particular for energy
splittings exceeding the optical phonon energy.
In general, decoherence in such systems may be dominated by pure
dephasing due to occupation-conserving phonon scattering
\cite{machnikowski06d}.  

Our findings seem to be consistent with the general features of
experimental observations. The 
size range where the relaxation is very
efficient (on picosecond time scales) is extremely narrow and does not
exceed a few Angstrom, which is 
comparable to the lattice constant of GaAs. This means that such an efficient
relaxation between the two lowest states in self-assembled quantum dot
molecules is a rather rare phenomenon which occurs only for very
finely tuned (accidentally or intentionally) dots and is unlikely to
be observed in a typical sample. Therefore, we conclude that
relaxation times on the order of at least hundreds of
picoseconds should be typical. The coupling between the dots decreases
exponentially with the distance between them which reduces the overlap
between the wave functions. Therefore, phonon-assisted tunneling for a
spontaneously formed pair of non-identical dots should become
inefficient as the spatial separation between between the
dots grows beyond 
a certain distance, as is indeed observed in experiments
\cite{heitz98,tackeuchi00,mazur05,reischle07}. In the case studied
here, that is, single-phonon relaxation between states separated by a
few meV in energy, the relaxation rates undergo oscillations as
functions of the geometrical parameters due to a
structured nature of the phonon reservoir and the resulting
interference effects. One should note, however, that most of the
available experimental data correspond to systems which much larger
energy splittings.

A more quantitative comparison is possible in the case of the
measurements presented in Ref.~\onlinecite{nakaoka06}. Here, electron
tunneling (that is, a transition between spatially direct and indirect
exciton states) has been studied for a QDM with a fixed 10~nm spacing and
energy level difference of a few meV, which corresponds more closely
to the physical situation of our model. Our
calculations for such parameter range yield transfer times in the range
of hundreds of picoseconds, which reasonably agrees with the measured
time of 0.5~ns (note that a slightly different material system was
used in that experiment and that some details of the system geometry
are not exactly known). It will be interested to include the electric
field in our model and to seek a closer correspondence with the
experiment, which is planned as a future work.

\begin{acknowledgments}
This work was partly supported by Grant No. N N202 1336 33 of the Polish
MNiSW.
\end{acknowledgments}

\appendix

\section{Form factors and Coulomb matrix elements}

In this appendix, we briefly summarize the method of calculating the
form factors and Coulomb matrix elements based on the wave functions
obtained within the variational multi-component envelope function
formalism using the simplification offered by a cylindrically symmetric system.

Using the identity
\begin{equation*}
\frac{1}{\vert \bm{r}-\bm{r}' \vert} = 
\frac{1}{(2 \pi)^{3}} \int d^{3} q\frac{4 \pi}{q^{2}} 
e^{i \bm{q}\cdot( \bm{r} - \bm{r}' )},
\end{equation*}
one writes the Coulomb matrix element $v_{ijkl}$ given by
Eq.~(\ref{v-ijkl}) in the form
\begin{equation}\label{v-ijkl-app}
v_{ijkl}=\frac{e^{2}}{(2 \pi)^{3} \varepsilon_{0} \varepsilon} \int
\frac{d^{3} \bm{q}}{q^{2}} \ff_{il}(\qq) \ff_{kj}^{*}(\qq), 
\end{equation}
where the form factors are given by Eq.~(\ref{FF}). 

We will use cylindrical coordinates for the vector
$\rr=[\rho\cos\phi',\rho\sin\phi',z]$ and
spherical coordinates for the vector
$\qq=[q_{\bot}\cos\phi,q_{\bot}\sin\phi,q_{z}]$, where 
$q_{\bot}=q\sin\theta$ and $q_{z}=q\cos\theta$.
For wave functions in the form given in Eq.~(\ref{ansatz}), one has
\begin{equation}\label{ff1}
\ff_{kj}(\qq) = 
\tilde{\ff}_{kj}(q_{\bot},q_{z}) 
i^{M_{j}-M_{k}}  e^{i  (M_{j}-M_{k})  \phi},
\end{equation}
where 
\begin{eqnarray}
\tilde{\ff}_{kj}(q_{\bot},q_{z}) &=& 
\sum_{\alpha,\beta}  \int_{0}^{\infty} \rho d \rho   
\chi_{\alpha \beta}(\rho,q_{z})  \varphi^{(k)}_{\alpha}(\rho)  
\varphi^{(j)}_{\beta}(\rho) \nonumber \\
&& \times\mathcal{J}_{M_{j}-M_{k}}(q_{\bot} \rho ).
\label{ff2}
\end{eqnarray}
In Eqs.~(\ref{ff1}) and (\ref{ff2}), we denoted the angular momenta of
the two states by $M_{k},M_{j}$, used the identity  
\begin{eqnarray*}
 \lefteqn{\frac{1}{2 \pi} \int_{0}^{2 \pi} d \phi'  e^{i [ (m'-m) \phi' +
   a \cos{(\phi' - \phi})] }  =}\\
&& \mathcal{J}_{m'-m}(a)  i^{m'-m} e^{i(m'-m)\phi}, 
\end{eqnarray*}
where $\mathcal{J}_{m}$ is the $m$-th Bessel function, and introduced
the quantities
\begin{equation*}
\chi_{\alpha \, \beta}(\rho,q_{z}) = \int_{-\infty}^{\infty} d z 
\chi_{\alpha}(\rho,z) e^{i q_{z} z}  \chi_{\beta}(\rho,z),
\end{equation*}
which are calculated by fast Fourier transform on the grid.

Substituting Eqs.~(\ref{ff1}) and (\ref{ff2}) into
Eq.~(\ref{v-ijkl-app}) and integrating over $\phi$ one finds
\begin{eqnarray*}
v_{ijkl}
& = & \delta_{M_{i}+M_{j},M_{k}+M_{l}} 
\frac{e^{2}}{(2 \pi)^{2} \epsilon_{0} \epsilon} 
\int_{0}^{\pi} d \theta\sin\theta
\int_{0}^{\infty} d q\\
&&  \times\tilde{\ff}_{il}(q \sin\theta,q \cos\theta) 
\tilde{\ff}_{kj}^{*}(q \sin\theta,q \cos\theta).
\end{eqnarray*}


\end{document}